\documentclass{PoS}

\usepackage{xspace}
\usepackage[utf8]{inputenc}

\title{Prototype operations of atmospheric calibration devices for the Cherenkov Telescope Array}

\ShortTitle{Operations of atmospheric calibration devices for CTA}

\author{\speaker{Jan Ebr}$^a$, Dušan Mandát$^a$, Miroslav Pech$^a$, Ladislav Chytka$^b$, Jakub Juryšek$^{a,b}$, Michael Prouza$^a$, Petr Janeček$^a$, Petr Trávníček$^a$, Jiří Blažek$^a$ and Markus Gaug$^c$ for the CTA consortium and Martin Mašek$^a$, Jiří Eliášek$^a$ and Sergey Karpov$^a$\\
\llap{$^a$}\textit{FZU -- Institute of Physics of the Czech Academy of Sciences, Czech Republic}\\
\llap{$^b$}\textit{Palacký University, Olomouc, Czech Republic}\\
\llap{$^c$}\textit{ Departament de Fisica, and CERES-IEEC, Universitat Autonoma de Barcelona, Spain}\\ 
E-mail: \email{ebr@fzu.cz}}

\abstract{The atmospheric monitoring devices for the planned calibration system of the Cherenkov Telescope Array (CTA) are undergoing intensive development, prototyping and testing. The All-Sky Cameras, the Sun/Moon Photometers and the FRAM telescopes have been gradually deployed at the future CTA sites with the primary goal of site characterization, simultaneously allowing the assessment of their operational reliability in realistic environmental conditions. All three devices have shown the ability to work smoothly in both the extreme dryness and the large temperature variations of the southern site as well as in the occasional adverse weather during winter months at the northern site. The target availability of 95\% of time has not yet been reached mostly due to minor hardware failures that have proven difficult to fix due to the remoteness of the installation in the absence of the future CTA infrastructure. The experience gathered during the prototype operations will contribute to the improved reliability of the final instruments. The Raman LIDARs, described in separate proceedings of this conference, and the infrared Ceilometer, ready for testing in Prague, will complement the set of atmospheric calibration devices in near future. The final operational procedures for the atmospheric calibration of the CTA during its operation are being finalized foreseeing the use of the All-sky Cameras and the Ceilometer for the monitoring of clouds over the whole sky and the LIDARs and FRAMs for precision measurements of the atmospheric transmission as a function of altitude and position within the field-of-view of the CTA array.}

\FullConference{36th International Cosmic Ray Conference -ICRC2019-\\
July 24th - August 1st, 2019\\
Madison, WI, U.S.A.}

\begin{document}

\begin{figure}
\centering
\includegraphics[width=0.99\textwidth]{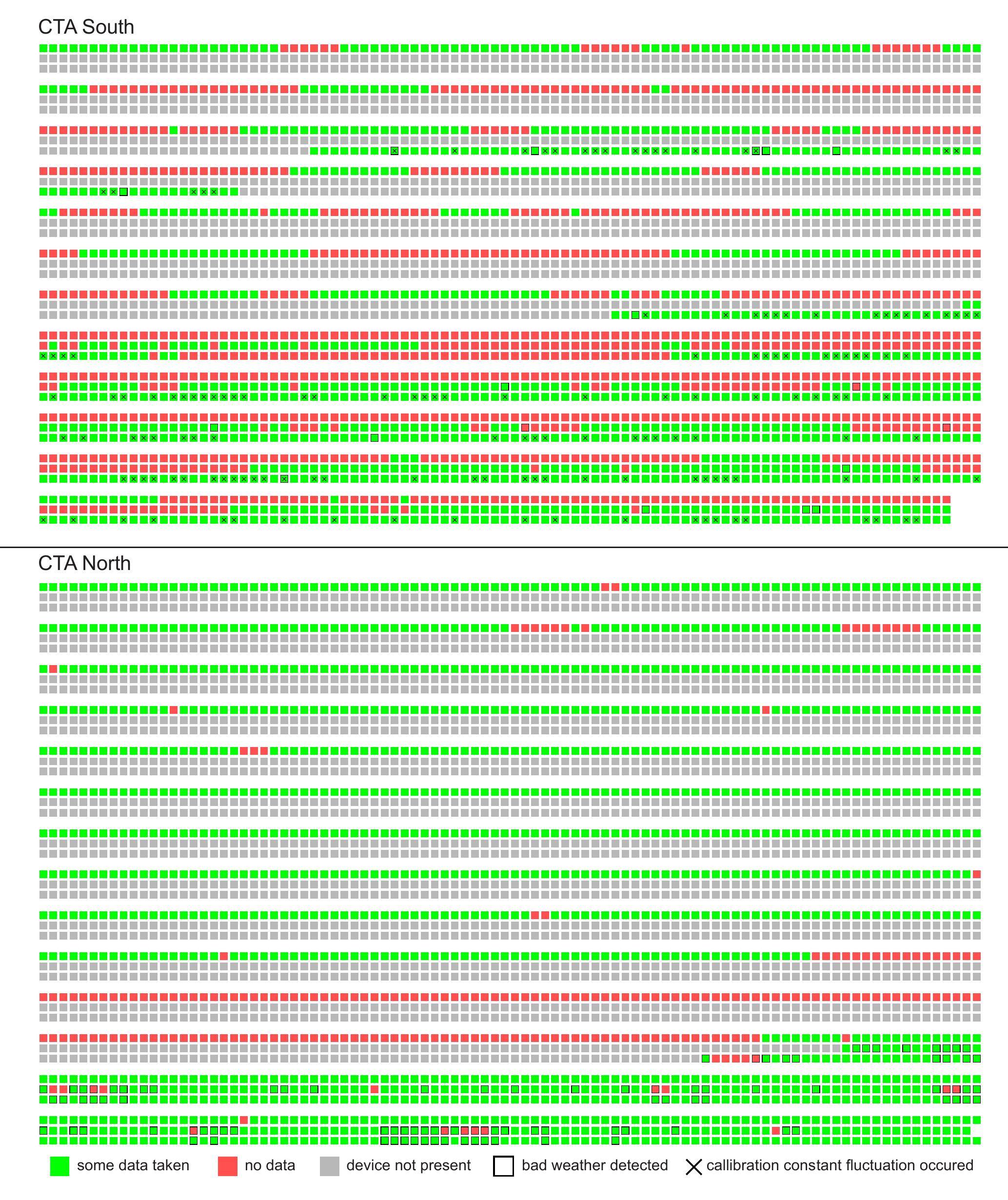}
\caption{\label{fig:uptime} Uptime of the prototype devices at the future CTA sites. In each series, the first line corresponds to the All-sky Camera, the second line to FRAM and the third line to the Sun/Moon Photometer. For CTA South the time period covered is 28/11/2015--26/12/2018, for CTA North it is 15/10/2015--22/05/2019. For each day, the color of the box indicates whether at least some data have been taken, or the instrument was not present at all. It is also shown when the device detectors reported bad weather (rain or high winds) incompatible with device operations, which could be a cause of missing data; in most cases however, the bad weather period did not cover the whole day. For the Sun/Moon Photometer also instances of fluctuating calibration constants are shown; whether the data from these days is recoverable is not clear at the moment. }
\end{figure}

\section{Experience with prototypes at future CTA sites}

Atmospheric monitoring devices, such as the All-sky Cameras (ASC), the Sun/Moon Photometers and the FRAM robotic telescopes have been, since 2015, gradually deployed on the future sites of the Cherenkov Telescope Array (CTA) -- the CTA South site near Cerro Armazones, Chile and the CTA North site on La Palma, Canary Islands, Spain \cite{CTA}. This serves two purposes: to characterize the conditions at the sites in preparation for the CTA operational phase (see \cite{sitechar} for results) and to test the operation of the devices in realistic conditions, improve their reliability and develop maintenance procedures, as well as to learn how to process the data obtained in the extremely clear atmosphere of the site. The prototype operation on CTA South was interrupted by an intrusion in 12/2018 when all the solar power equipment was stolen and any remaining hardware had to be temporarily moved to storage. The devices will be installed again at a safer location and the operation will resume in 09/2019; the operation at CTA North is ongoing.

The uptime of the devices on both sites is summarized in Fig~\ref{fig:uptime}, which indicates, for every day, whether at least some data have been taken by the given instrument. The causes for the downtimes observed and their impact on the future operation of the instrument within CTA are discussed for each device individually. 

\subsection{All-sky Cameras}

The All-sky Camera system \cite{ASC} consists of G2-4000 CCD camera by Moravian Instruments equipped with photometric Johnson and UV filters and Sigma 4.5 mm f/2.8 EX DC fish-eye lens connected to a data acquisition computer. Initially the computer was an Intel Atom-based single board computer running embedded Windows XP system and Matlab code for data acquisition. Due to several failures of the computers running with ASCs at Pierre Auger Observatory and at CTA South and unavailability of replacements, the acquisition system was replaced in 2017 by Raspberry Pi 3 single board computer running Linux and Python code.

The ASC at the CTA North site was collecting data used to determine the cloudiness on the site smoothly most of the time. Two water leaks into the ASC housing occurred in April 2018 and April 2019. The issues were caused by a degraded sealant and a better silicone sealant was identified as a replacement. The operation of the ASC at the CTA South site was more problematic due to the off-grid operation of the device depending on solar power. Around April 2016 insufficient power in the single installed battery was detected and another battery was added, which stabilized the power for a while. Later in 2016, the computer died and was replaced after few months. Issues with power started to appear again in 2017 and eventually the charger failed in September 2017. The charger was replaced in July 2018 and the Raspberry Pi was installed at the same time. However, the batteries reached the end of their lifespan and could not be replaced at the time, therefore only very limited operation was possible since. Overall, even the few failures caused long downtimes due to the remote operation and essentially no availability of local staff with spare parts. This will clearly not be the case during the CTA Observatory operation.

\subsection{FRAM telescopes}

The FRAM \cite{FRAM} is a robotic telescope, using an off-the-shelf astronomical equatorial mount (Paramount MYT for CTA South and 10Micron GM2000HPS for CTA North) to guide a Zeiss 135/2 lens attached to a Moravian Instruments G4-16000 large-format CCD camera, for the purpose of the determination of atmospheric extinction using wide-field stellar photometry \cite{FRAM}. A significant part of the downtime observed at CTA South was due to interruptions of internet connection to the remote site, as in the early stages, the operating software was not suitable for completely unsupervised operations -- this has been gradually improved over time to the point where autonomous operation is now possible for many days. Two major hardware incidents were recorded with the FRAM at CTA South. In 09/2017, the protective roof did not fully close, due to a leak in the hydraulic system, which became loose over time -- this seems to be due to "settling" of the new installation and will be prevented by checking any newly installed, or re-hosed FRAM once, one month after installation. The cooling of the CCD camera stopped working in 06/2018 due to a damage of the camera mainboard. The vendor determined humidity as the cause of the issue and suggested shortening the maintenance intervals of the camera desiccant to half a year to prevent future issues. Both failures could be easily solved within a day if there was technical staff and spare parts on site, as foreseen during CTA operations. The CTA North FRAM experienced only small interruptions as the reliable connection to the site provided by the MAGIC collaboration has been very helpful. At least some downtime can be explained with bad weather and the rest mainly due to tweaking of the drivers for the previously not used 10Micron mount. However, determining whether bad weather is fully responsible for a whole night of downtime is difficult from the system logs -- upon discovering this issue, we have started keeping a daily hand-written log of operations for better assessment of prototype reliability.

The operation of the FRAM telescopes under extremely clear conditions has also aided the development of improved methods of determination of the vertical aerosol optical depth (VAOD) from FRAM data. The mostly low and almost constant amount of aerosols at both sites allows detection of instrumental artifacts, such as the population of "outliers" with VAOD about 0.05 higher than surrounding points, that were traced to a rare error when the telescopes moves during a part of the exposure. In a similar vein, the clarity of the data helped to find the source of the unexpected correlation of the data with the phase of the Moon, as reported in \cite{FRAM2}.

\subsection{Sun/Moon Photometer}

The Sun/Moon photometer CE318-T is a commercial device which measures atmospheric extinction using the Sun or the Moon as a light source. During the time of its deployment on both sites, it has worked reliably, with the exception of an incident on CTA South in 10/2017 when it ran out of space in internal storage to record data. This problem can be circumvented by connecting the device to a computer, however this has proven to be problematic, because in case of problems with such a connection, data may be lost altogether. As storing data only within the device does not allow monitoring of its status, the CTA North Photometer has been endowed with a module for data connection over cellular network, which however is yet to be connected to the network. 

\begin{figure}
\centering
\includegraphics[width=0.99\textwidth]{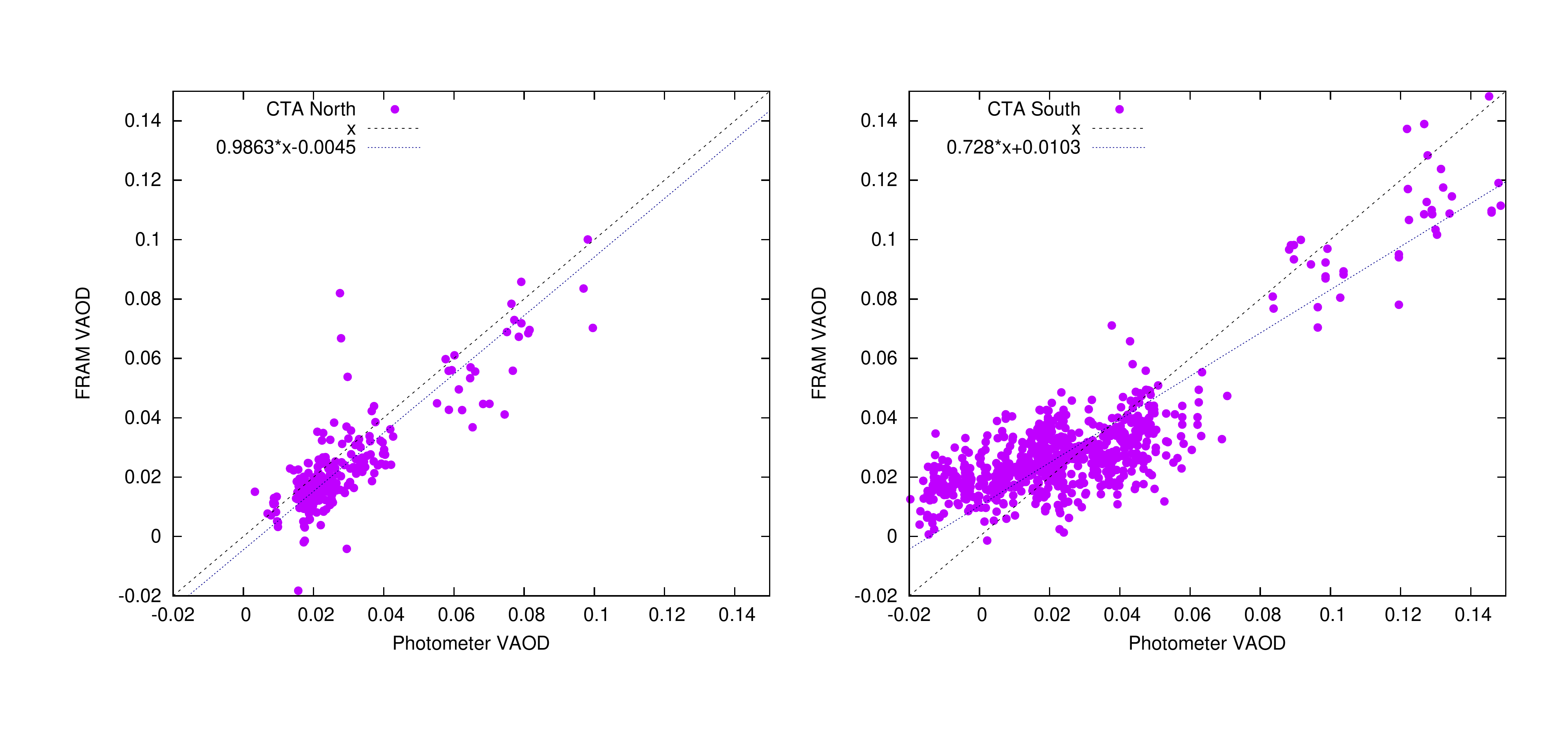}
\caption{\label{fig:framphot} Comparison of FRAM and Moon Photometer VAOD measurements taken within 15 minutes from each other. Cuts on calibration constant stability and Moon phase were applied to CTA South Photometer data. In each panel, a linear fit is plotted.}
\end{figure}

Despite the excellent track record of operation, the actual availability of high-quality data from CTA South is limited by unexpected fluctuations in the calibration constant of the device. This value is expected to be stable with gradual decay over time, but it varies significantly during some periods, rendering the data occasionally nonphysical (negative VAODs occur). It is not yet clear whether the affected data can be recovered (it will be discussed with the vendor). The difference in the behavior between the two Photometers can be seen when compared to FRAM data (Fig.~\ref{fig:framphot}), as the CTA South Photometer shows significant deviations even after stringent cuts are applied on the quality of the calibration and on the Moon phase (as the data quality is worst for small lunar illuminations).

\section{Operation schemes for the future CTA Observatory}

In parallel to the progress in hardware and data analysis, the concepts for integration of these devices into the daily operations of the future CTA Observatory are being actively developed. The aerosol information provided by the FRAM telescopes in concert with Raman LIDARs to be installed at the sites will be used for data analysis and corrections, both \textit{online} and \textit{offline}. The ASCs will be paired with Ceilometers to provide 3D information on clouds for the purpose of intelligent scheduling.

\subsection{Atmospheric calibration}

In order to achieve the required level of accuracy for the \textit{offline} analysis, contemporaneous and only marginally simplified aerosol profiles must be
used to simulate the CTA's instrument response function (IRF). Only for the \textit{online} analysis, faster, though less accurate, algorithms may be used which corrects the
effective areas and energy biases only (see e.g.~\cite{Fruck:2015}).

\begin{figure}[t]
\centering
\includegraphics[width=0.99\textwidth]{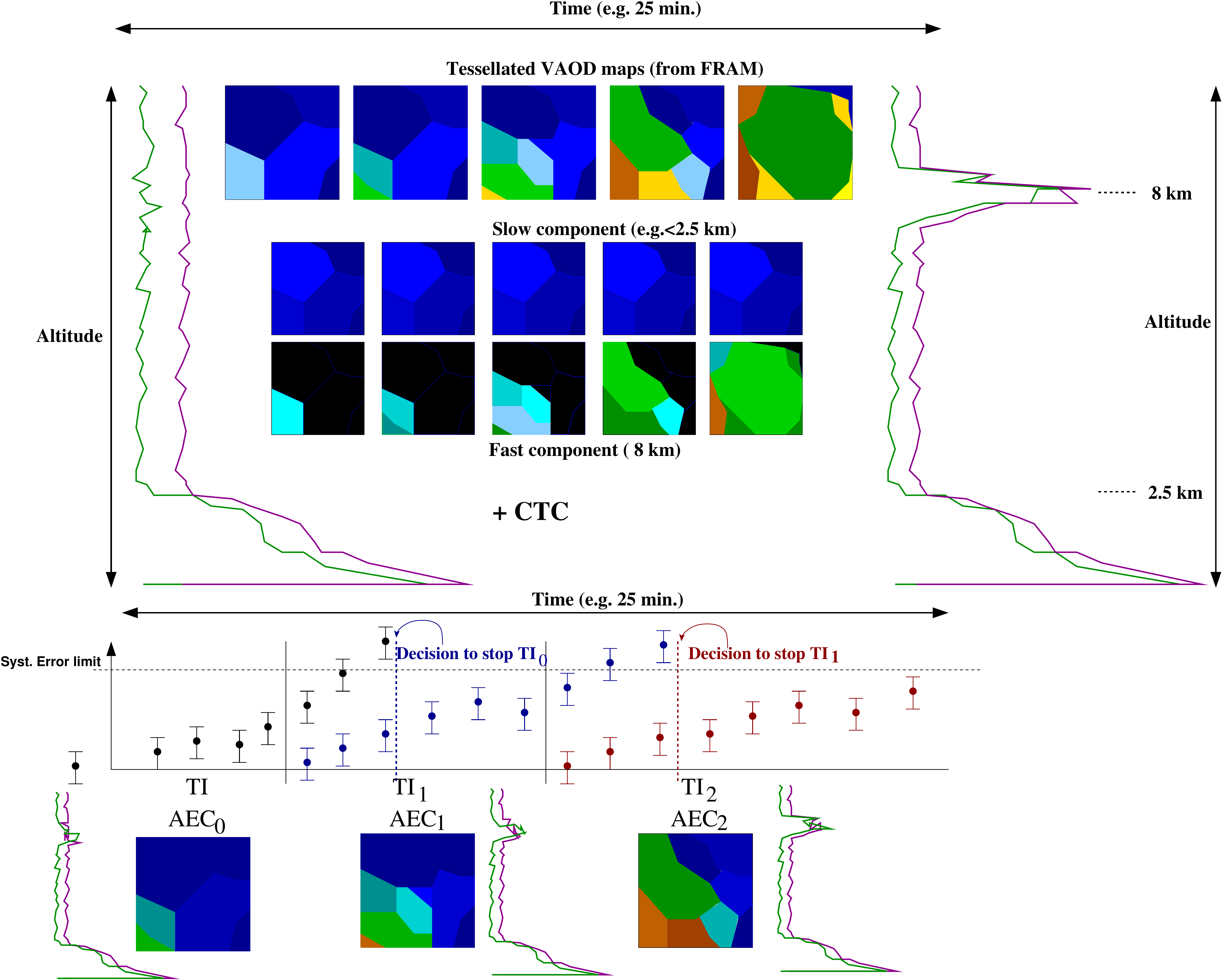}
\caption{\label{fig:atmocorrections} Scheme for the determination of new GTIs and new MC simulated IRFs.}
\end{figure}

Figure~\ref{fig:atmocorrections} highlights then the procedure to obtain average instrument response functions over a time interval within which the systematic error due to
simplifications of the profile remains acceptable: 

\begin{enumerate}
\item Measurements of the full profile are taken with the Raman LIDAR \cite{Raman} before and after an observation block.
Between these, the FRAM follows the field-of-view of the CTA array and measures the integral AOD.
At the same time, the CTA telescopes register the trigger rates.
\item The individual integral AOD measurements of detected stars are combined within suitable patches to produce AOD maps (e.g. using a Voronoi tesselation technique~\cite{Janecek:2017}).
A possible stratospheric aerosol contribution must be directly subtracted at this point.
The CTC (Cherenkov transparency coefficient) is calculated from the telescope trigger rates and the optical throughput calibration~\cite{muons},
according to the prescriptions outlined in~\cite{CTC}.
\item The AOD maps are interpolated in time. Interlaced Raman LIDAR measurements may serve to divide the interpolated AODs
in vertical bins and to re-calibrate its integral. 
\item These ``aerosol extinction hypercubes'' (with altitude, wavelength and time as additional dimensions) are then split into a 
slow component (due quasi-stable aerosol layers, like the boundary layer or sometimes even clouds), and 
a fast one, which changes throughout a science observation run. 
\item The expected shifts in CTA performance from a given extinction hypercube are continuously confronted with those 
of currently used (or foreseen) MC generated IRFs, 
and a series of systematic errors is calculated and confronted with the CTA requirements.
\item In case one of these systematic errors exceeds the allowed limit,
a new ``Good Time Interval'' (GTI) is communicated to CTA, 
together with a suitable start time, and an averaged extinction hypercube, to be used for a new MC simulation of the atmosphere.
The average extinction cube (AEC) contains the vertical profile in one dimension, the wavelength dependency in a second dimension, and the aerosol extinction coefficient as a result of both.
The AEC must be quality checked and confronted with alternative procedures, like the CTC.
\end{enumerate}

The critical part of this procedure is a robust estimate of the systematic error. 
For this purpose, MC simulation studies are required to determine the impact of aerosols and clouds at different altitudes, and cirrus clouds covering parts of the FOV of a Cherenkov camera
on the angular and energy resolution and bias, if a uniform coverage of the CTA cameras by aerosols has been assumed in the simulations.
Previous studies of the typical morphology of clouds and evolution time scales at the CTA sites are needed as input for these simulations.

\subsection{Intelligent target selection}

For the intelligent target selection, a robust prediction of the environmental conditions across the full observable sky,
for at least the duration of the next scheduling block (i.e. $\sim$20~min) is
necessary, but the accuracy of that prediction is of a less stringent requirement.

This leads to the following operation scheme:

\begin{enumerate}
\item The ASC takes a picture of the full sky and converts the result in an atmospheric extinction map (AEM).
\item A cloud recognition program applies a threshold to the AEM and decomposes the result in ellipses~(see e.g.~\cite{adam}),
yielding the ASC Clouds List (CL). 
\item 
The Ceilometer receives the full list of possible observation targets in the short-term scheduler and picks those which are close enough to a cloud in the CL as to get possibly affected by it.
After discarding those clouds that remain not associated with CTA observation targets, the Reduced ASC Clouds List (RCL) is obtained.
The Ceilometer points to the center of each cloud in the RCL and takes an extinction profile.
The mean altitude of each cloud is added to the RCL. 
\item A now-cast of wind speed and direction at the relevant cloud altitudes is obtained from Global Data Assimilation Systems~\cite{munar} and added to the RCL.
Using the RCL and historical knowledge, each cloud in the RCL gets propagated in time for a duration corresponding to next scheduling block,
and a probability is calculated for each associated CTA observation target to get covered by the cloud.
\item In case the probability lies above a certain threshold, the expected observation parameters for the next scheduling block (reduction of accessible energy range, degradation of energy and pointing resolution and effective areas) are calculated.
\item The list of targets with associated cloud coverage probabilities and performance degradation parameters is returned to the CTA scheduler for evaluation,
before the next scheduling block is executed.
\end{enumerate}

\section*{Acknowledgements}

We gratefully acknowledge financial support from the agencies and
organizations listed here: \href{http://www.cta-observatory.org/consortium\_acknowledgments}{www.cta-observatory.org/consortium\_acknowledgments}, in particular by Ministry of Education, Youth and Sports of the Czech Republic (MEYS) under the projects MEYS LM2015046, LTT17006 and 
EU/MEYS CZ.02.1.01/0.0/0.0/16\_013/0001403 and European Structural and Investment Fund and MEYS (Project CoGraDS -- CZ.02.1.01/0.0/0.0/15\_003/0000437). The installation of the devices on La Palma would not be possible without the support from the MAGIC Collaboration. This work was conducted in the context of the CTA Consortium.

\end{document}